\documentclass[12pt]{article}

\usepackage{epsfig,epstopdf,latexsym,amsfonts,amsmath,amsthm,amssymb,amsbsy,multirow,slashed,wasysym,textcomp,wrapfig,graphicx,psfrag,booktabs,bbm,comment}

\usepackage[toc]{appendix}

\usepackage{color}
\usepackage{datetime}
\usepackage[
      colorlinks=false,
      linkcolor=darkblue,  
      urlcolor=blue,    
      filecolor=blue,     
      citecolor=red,
linktocpage=true,
      pdfstartview=FitV,
      bookmarksopen=true    
      ]{hyperref}

\ifpdf
\fi

\numberwithin{equation}{section}

\renewcommand{\theequation}{\arabic{section}.\arabic{equation}}

\def\coeff#1#2{\relax{\textstyle {#1 \over #2}}\displaystyle}

\def\IR{\mathbb{R}}
\def\ZZ{\mathbb{Z}}

\def\cB{{\cal B}}

\def\cF{{\cal F}}

\definecolor{cardinal}{rgb}{0.6,0,0}
\definecolor{darkgreen}{rgb}{0,0.5,0}
\definecolor{golden}{rgb}{0.92, 0.7, 0}
\definecolor{midnight}{rgb}{0, 0, 0.5}
\definecolor{darkblue}{rgb}{0.2, 0, 0.8}

\begin{document}

\begin{titlepage}

 \begin{flushright}
IPhT-T13/036
 \end{flushright}

\bigskip
\bigskip

\centerline{\Large \bf Almost BPS but still not renormalized}

\medskip
\bigskip
\bigskip
\centerline{{\bf Iosif Bena$^1$, Andrea Puhm$^1$, Orestis Vasilakis$^2$ and Nicholas P. Warner$^{1,2,3}$}}
\centerline{{\bf }}
\bigskip
\centerline{$^1$ Institut de Physique Th\'eorique, }
\centerline{CEA Saclay, CNRS-URA 2306, 91191 Gif sur Yvette, France}
\bigskip
\centerline{$^2$ Department of Physics and Astronomy}
\centerline{University of Southern California} \centerline{Los
Angeles, CA 90089, USA}
\bigskip
\centerline{$^3$ Institut des Hautes Etudes Scientifiques}
\centerline{ Le Bois-Marie, 35 route de Chartres}
\centerline{Bures-sur-Yvette, 91440, France}
\bigskip
\bigskip
\centerline{{\rm iosif.bena@cea.fr, andrea.puhm@cea.fr,  vasilaki@usc.edu, ~warner@usc.edu} }
\bigskip
\bigskip

\begin{abstract}

\noindent A key feature of BPS multi-center solutions is that the equations
  controlling the positions of these centers are not renormalized as
  one goes from weak to strong coupling.  In particular, this means
  that brane probes can capture the same information as the fully
  back-reacted supergravity solution. We investigate this
  non-renormalization property for non-supersymmetric, extremal
  ``almost-BPS'' solutions at intermediate coupling when one of the
  centers is considered as a probe in the background created by the
  other centers. We find that despite the lack of supersymmetry, the
  probe action reproduces exactly the equations underlying the fully
  back-reacted solution, which indicates that these equations also do
  not receive quantum corrections. In the course of our investigation
  we uncover the relation between the charge parameters of almost-BPS
  supergravity solutions and their quantized charges, which solves an
  old puzzle about the quantization of the charges of almost-BPS
  solutions.

\end{abstract}

\end{titlepage}

\tableofcontents
\newpage
\section{Introduction}

The physics of multi-center four-dimensional BPS solutions and of
their five-dimensional counterparts
\cite{Denef:2000nb,Denef:2002ru,Bates:2003vx,Gauntlett:2004qy,Bena:2005va,Berglund:2005vb,Saxena:2005uk}
has been one of the keys that could unlock longstanding mysteries of
black hole physics, such as the information paradox and the
microscopic origin of the black-hole entropy. These solutions yield
the easiest-to-construct black-hole microstate geometries
\cite{Bena:2006kb,Bena:2007qc} and can be used to study the wall-crossing behavior
of black-hole partition functions
\cite{Denef:2007vg, Manschot:2010qz}.  They also provide the
best-known examples of entropy enigmas
\cite{Gauntlett:2004wh,Denef:2007vg,Bena:2011zw}.
 
Another key feature of BPS, multi-center solutions is that the equations
controlling the positions of these centers, also known as the bubble
equations, are not renormalized as one goes from weak to strong
effective coupling. At weak coupling, these configurations are
described by a supersymmetric quiver quantum mechanics (QQM)
\cite{Denef:2002ru}, and the equations determining the vevs of the QQM
Coulomb-branch fields are the same as those determining the
inter-center distances of the fully-back-reacted supergravity
solution. This remarkable non-renormalization property has allowed one
to compute the symplectic form, quantize the moduli space of supergravity
solutions from the QQM perspective \cite{deBoer:2008zn}\footnote{This in turn has confirmed that
quantum effects can wipe out macroscopically large regions of certain smooth
low-curvature solutions \cite{Bena:2007qc,deBoer:2008zn}.} and has given a 
clear mapping of some of the microscopic black-hole degrees of
freedom to horizonless solutions that exist in the same regime of the
moduli space as the classical black hole \cite{Bena:2012hf,Lee:2012sc,Manschot:2012rx}.

For BPS solutions one can also test this non-renormalization of the
bubble equations by considering multi-center solutions in an
intermediate region of the effective coupling, where some of the
branes have back-reacted while others are treated as probes. More
precisely, one can place a supertube in a multi-center solution and
examine its supersymmetric minima. The equations that determine the
positions of these minima can then be shown to be identical to the
bubble equations \cite{Bena:2008dw}. Hence, one can recover the BPS
bubble equations both at small effective coupling from the QQM, at
intermediate coupling from the supertube DBI action, and at large
effective coupling by asking that the fully back-reacted solution has
no closed timelike curves.

Since the non-renormalization of the bubble equations is established
by invoking quantities protected by supersymmetry, one might naively
expect that the beautiful pieces of physics that are protected from
renormalization in BPS systems would no longer survive in
non-supersymmetric, multi-center solutions or in non-supersymmetric
fuzzballs. The purpose of this paper is to demonstrate that, on the
contrary, in certain classes of non-supersymmetric multi-center
solutions - the so-called almost-BPS solutions \cite{Goldstein:2008fq,Bena:2009ev,Bena:2009en} - the bubble equations
are also protected when one goes from intermediate to strong coupling.
This suggests that there are no quantum corrections despite the lack
of supersymmetry.

This result is quite surprising because there is neither supersymmetry
nor any other underlying symmetry that would prevent the bubble
equations from receiving corrections and because the bubble equations
of almost-BPS solutions contain complicated cubic combinations of the
inter-center distances, whose coefficients do not {\it a priori}
appear to be related to anything one can define in a quiver quantum
mechanics. However, in retrospect, this result may not look so surprising: 
Emparan and Horowitz have shown in \cite{Emparan:2006it}
that the entropy of certain extremal non-supersymmetric black holes
can be calculated at weak effective coupling, and does not change as
one increases this coupling. Since a certain subclass of almost-BPS
solutions can be dualized to the black hole considered in
\cite{Emparan:2006it}, it seems plausible that whatever principle
protects the quantum states calculated at weak coupling from being
uplifted and disappearing at strong coupling also protects the bubble
equations from receiving quantum corrections\footnote{It is also
interesting to note that whatever underlying principle protects the
states and bubble equations of almost-BPS system from being uplifted
by quantum corrections can be even stronger than supersymmetry,
which, for example, does not protect the degeneracy of the D1-D5-P
system in the non-Cardy regime \cite{Bena:2011zw}.}.

To establish that the almost-BPS bubble equations are not changing
with the coupling, we consider an intermediate-coupling configuration
where all the centers, except one, have back-reacted into an almost-BPS
supergravity solution.  One can then probe this solution with a
supertube and use the (Dirac-Born-Infeld and Wess-Zumino) action to
determine the equilibrium position of the probe.  If one considers the
back-reaction of the probe then the supertube would become another
center of the supergravity solution and its location is fixed by the
bubble equations, which arise from requiring that there be no closed
time-like curves.  The comparison of these two physically distinct
conditions, one in field theory on the brane and the other from its
gravity dual, yields the test of non-renormalization between the
intermediate and strong coupling regimes.  At first glance, the
equations look nothing like each other: both differ from the BPS
bubble equations by extra cubic terms, but these cubic terms are not
the same. However, the story is a bit more complicated: Unlike the BPS
solution, the quantized electric charge of the supertube DBI action
and the electric charge parameters in the supergravity harmonic
functions are not the same.  One can then ask whether there exists a
relation between the DBI and the supergravity electric charges of the
supertube that maps one set of bubble equations onto the other. One of
the primary results of this paper is to prove that such a relation
exists, and write it down for the most general known almost-BPS
solution in a single-center Taub-NUT space, involving an arbitrary
collection of concentric supertubes and black rings.

The implications of this relation for the physics of almost-BPS
solutions are quite significant.  First, in a general supergravity
solution that contains several supertubes, the quantized charges of
these supertubes are not the obvious coefficients in the supergravity
harmonic functions. The supergravity parameters are related to the
quantized charges by certain shifts that come from dipole-dipole
interaction and depend on the location of all the other centers of the
solution. An interesting features of these shifts is that they only
depend on the dipole charges and positions of the centers that lie
between location of the charge  in question and the 
(supersymmetry-breaking) center of the Taub-NUT space.

For almost-BPS black rings the effect of the charge shift is more
subtle because the black rings have more than one dipole charge, and
hence the formulae that give the dipole-dipole contribution to the
charge appear to be degenerate. However one can assemble a black ring
by bringing together three supertubes with different kinds of dipole
charges, and this will allow us to calculate the shifts between the
black-ring supergravity and the quantized charges.

The second significant implication has to do with charge quantization,
and solves a longstanding puzzle of the physics of almost-BPS black
rings and of multi-center almost-BPS solutions. As one can see from
\cite{Bena:2009en} or from \cite{Vasilakis:2011ki}, the (quantized)
asymptotic charge of such solutions is equal to the sum of the
supergravity charges of the centers plus an extra dipole-dipole term
that depends on the positions of the centers and the moduli of the
solution. If the supergravity charges were equal to the quantized
ones, this would have implied that the moduli-dependent contribution
to the charges are also quantized, and hence multi-center almost-BPS
solutions could only exist on special codimension-three slices of the
moduli space (where the moduli-dependent contributions are integers).
This would have been quite puzzling. The results presented here show 
that this does not happen.  Upon using
our formulas that relate the supergravity and the quantized charges,
one finds that the asymptotic charge of the almost-BPS black ring is
equal to the sum of the quantized charges of its centers, and hence
the almost-BPS black ring as well as other multi-center almost-BPS
solutions exists for any values of the moduli. If one keeps the
quantized charges fixed and changes the moduli, the quantities that change
are simply the supergravity charge parameters of the centers. Our result also
implies that the $E_{7(7)}$ quartic invariant of the almost-BPS black
ring of \cite{Bena:2009ev} needs to be rewritten in terms of the
quantized charges, and also yields the relation between the
``quantized'' angular momentum of this black ring and the
angular-momentum parameter that appears in the corresponding
almost-BPS harmonic function.

A third, perhaps more unexpected consequence of our result is that if
one probes a certain supergravity solution with a supertube and finds,
say, two minima, one where the supertube is at the exterior of the
other supertubes and one where it is at an intermediate position,
these two minima do not correspond to two vacua in the same vacuum
manifold of the bubble equations.  Indeed, one can relate both minima
of the supertube potential to almost-BPS supergravity solutions, and
then use our recipe to compute the quantized charges of the centers of
these two solutions. Since the shifts of the charges of the centers
depend on the position of the tube, the quantized charges of the
centers of the two resulting solutions will not be the same, and hence
these solutions do not describe different arrangements of the same
supertubes, and thus cannot be related by moving in the moduli space
of solutions of a certain set of bubble equations. Instead they live
in different superselection sectors.

In Section 2 we review the structure of BPS and almost-BPS
supergravity solutions and in Section 3 we calculate and examine the
action of supertube probes in these solutions. We show that one can
reproduce the supergravity bubble equation of the outermost supertube
by considering this supertube as a probe in the fully back-reacted
solution formed by the other supertubes and find the equation that
relates the supergravity charge parameters and the quantized charges
of the probe. We then give a recipe to read off the quantized
supertube charges in a general multi-center almost-BPS solution. In
Section 4 we then show that one can also recover the bubble equations
of all the other supertube centers by examining the minima of probe
supertubes and relating their supergravity charges to their quantized
charges. This demonstrates that all the supergravity data of a
multi-center solution can be recovered from the action of supertube
probes, and hence this data is not renormalized as one goes from weak
to strong effective coupling. We also discuss in more detail the
physics behind the shift needed to relate supergravity and quantized
charges. Section 5 contains our conclusions and a discussion of
further issues arising from this work.  Some of the technical details of this 
paper have been relegated to an appendix.

\section{The supergravity solutions }

\subsection{BPS and almost-BPS equations}

The simplest way to describe the solutions of interest is to work in M-theory, with the metric
\begin{multline}
ds_{11}^2  = - \left( Z_1 Z_2  Z_3 \right)^{-{2 \over 3}} (dt+k)^2
+ \left( Z_1 Z_2 Z_3\right)^{1 \over 3} \, ds_4^2  \\+ \left(Z_2
Z_3 Z_1^{-2}  \right)^{1 \over 3} (dx_5^2+dx_6^2) + \left( Z_1 Z_3
Z_2^{-2} \right)^{1 \over 3} (dx_7^2+dx_8^2)  + \left(Z_1 Z_2
Z_3^{-2} \right)^{1 \over 3} (dx_9^2+dx_{10}^2) \,,
\label{11Dmetric}
\end{multline}
where $ds_4^2$ is a four-dimensional hyper-K\"{a}hler metric.  The three-form potential is given by: 
\begin{equation}
\mathcal{A} = A^{(1)}\wedge dx_5 \wedge dx_6 + A^{(2)}\wedge dx_7
\wedge dx_8 + A^{(3)}\wedge dx_9 \wedge dx_{10} \,,
\label{11Dthreeform} 
\end{equation}
where the Maxwell fields are required to obey the ``floating brane Ansatz''  \cite{Bena:2009fi}:
\begin{equation}
A^I   ~=~  - \varepsilon\, Z_I^{-1}\, (dt +k) + B^{(I)}  \,,
\label{AAnsatz}
\end{equation}
and where $\varepsilon\ = \pm 1$ and $ B^{(I)}$ is a ``magnetic'' vector potential on the base, $ds_4$.  We further define the field strengths:
\begin{equation}
\Theta^{(I)}    ~\equiv~  d B^{(I)} \,.
\end{equation}

It was shown in  \cite{Bena:2009fi} that if one chooses the base to be merely Ricci-flat and solves the linear system of equations (for a fixed choice of $\varepsilon = \pm 1$): 
\begin{eqnarray}
\Theta^{(I)} &=&  \varepsilon *_4 \Theta^{(I)} \,,  \label{BPSeqna} \\  
\widehat \nabla^2 Z_I &=&   \coeff{1}{2} \, \varepsilon \, C_{IJK}   *_4 [\Theta^{(J)} \wedge \Theta^{(K)}] \,,\label{BPSeqnb} \\
 d k ~+~\varepsilon  *_4 d k  &=&   \varepsilon \,Z_I \Theta_I  \label{BPSeqnc}     \,,
\end{eqnarray}
then one obtains solutions to the complete equations of motion of
M-theory.  In particular one obtains a BPS solution
\cite{Bena:2004de,Gutowski:2004yv,Gauntlett:2002nw} by requiring that
the duality structure of the $\Theta^{(I)}$ matches that of the
Riemann tensor of the base.  That is, one obtains BPS solutions by
taking $\varepsilon = +1$ and choosing the base metric to be self-dual
hyper-K\"ahler, or by taking $\varepsilon = -1$ and choosing the base
metric to be anti-self-dual hyper-K\"ahler.  Almost-BPS solutions
\cite{Goldstein:2008fq,Bena:2009ev,Bena:2009en} are also obtained by
using hyper-K\"ahler base metrics but supersymmetry is broken by
mismatching the duality structure of the $\Theta^{(I)}$ and that of
the Riemann tensor of the base.

Here we will work with the Taub-NUT metric: 
\begin{equation}
ds^2_4 ~=~  V^{-1}(d\psi + A)^2 ~+~  V \, (dx^2 + dy^2 +   dz^2)     \,,
\label{TNmet}
\end{equation}
with 
\begin{equation}
 V ~=~  h ~+~ {q \over r}   \,,
\label{TNVdefn}
\end{equation}
where $r^2 \equiv \vec y \cdot \vec y$ with  $\vec y \equiv (x,y,z)$. We will also frequently use polar coordinates:  $x = r \sin \theta \cos \phi$, $y = r \sin \theta \sin \phi$ and $z = r \cos \theta$.   The metric (\ref{TNmet}) is hyper-K\"ahler if 
\begin{equation}
 \vec \nabla V ~=~ \pm  \vec \nabla \times \vec A   \,,
\label{VAreln}
\end{equation}
where  $\nabla$ denotes the flat derivative of $\IR^3$. 
For (\ref{TNVdefn}) we get
\begin{equation}
A =~ \pm q \, \cos \theta \, d \phi   \,,
\label{Aform}
\end{equation}
where we chose the integration constant to vanish.
%
%
We use the orientation with $\epsilon_{1234} =+1$ and then choosing the positive sign in (\ref{VAreln}) results in a the Riemann tensor that is self-dual while choosing the negative sign makes the Riemann tensor  anti-self-dual.  

Henceforth within this paper, we will follow the conventions of \cite{Vasilakis:2011ki} and take 
\begin{equation}
\varepsilon~=~ +1   \,,
\label{epschc}
\end{equation}
and then the BPS solutions and almost-BPS solutions correspond to choosing the $+$ or $-$ sign, respectively,  in  (\ref{VAreln}) and (\ref{Aform}).
 
\subsection{BPS and almost-BPS solutions}
\label{Sect:Sols}

We now  summarize, in some detail, the solutions presented in \cite{Vasilakis:2011ki}.  Since we are using the conventions (\ref{epschc}), there will be some sign differences compared to \cite{Vasilakis:2011ki}.

On a Gibbons-Hawking (GH) base the vector potentials, $B^{(I)}_\pm$  in (\ref{AAnsatz}) are  given by:
\begin{align} \label{Bterms}
& B^{(I)}_\pm   ~=~ P^{(I)}_\pm  (d\psi + A) ~+~ \vec \xi^{(I)}_\pm \cdot d \vec y \,,  \qquad P^{(I)}_+ ~ =~ V^{-1} \, K^I _+ \,;   \quad  P^{(I)}_- ~=~ K^I_-  \,,\\ 
& \nabla^2 K^I_\pm   ~=~0\,, \qquad    \vec \nabla \times \vec \xi^{(I)}_+  ~=~ - \vec\nabla K^I_+   \,, \qquad \vec \nabla \times \vec \xi^{(I)}_-  ~=~  K^I_-  \vec\nabla V  -  V \vec\nabla K^I_- \,,
\end{align}
where the $\pm$ corresponds to the choice in  (\ref{VAreln}) and hence to BPS or almost-BPS respectively.  

A ``type I'' supertube, $I=1,2,3$, has a singular magnetic source for $B^{(I)}$ and singular electric sources for $Z_J$ and $Z_K$, where $I,J,K$ are all distinct.   As in \cite{Vasilakis:2011ki},  we study an  axisymmetric supertube configuration with one supertube of each type on the positive $z$-axis and thus we take 
harmonic functions: 
\begin{equation}
K^I_\pm  ~=~ \frac{k_I^{\pm} }{r_I}  \,,
\label{Kfns}
\end{equation}
where
\begin{equation}
r_J~\equiv~  \sqrt{x^2 + y^2 + (z-a_J)^2}\,.
\label{ridefn}
\end{equation}
Without loss of generality we will assume that 
\begin{equation}
0<a_1 ~<~ a_2  ~<~a_3 \,.
\label{ajorder}
\end{equation}

We then have
\begin{equation}
 \xi^{(I)}_+   ~=~- k_I^+  \, \frac{(r \cos \theta  - a_I)}{r_I}\, d \phi  \,, \qquad  \xi^{(I)}_-   ~=~ - \frac{k_I^- }{a_I \, r_I}\,  \big[\, q \, (r   - a_I \cos \theta) +   h \, a_I \,   (r \cos \theta  - a_I) \, \big]\, d \phi   \,,
\label{xires}
\end{equation}
and we may then write
\begin{align}
 B^{(I)}_+    &~=~   \frac{K^I_+ }{V}\, \Big[\,  d \psi     + \frac{q\,  a_I }{r}\,  d \phi  -    h\, (r \cos \theta - \, a_I )\,  d \phi    \,  \Big] \,, \\
    B^{(I)}_-   & ~=~   \frac{k_I^- }{a_I \, r_I}\, \Big[\, a_I \, d \psi    -   q \,  r  \, d \phi -  h\,a_I \,   (r \cos \theta  - a_I)\, d \phi \,  \Big]    \,.
\label{BIres}
\end{align}

One can measure the local dipole strength by taking the integral of $\Theta^{(I)}$ over a small sphere, $S_\epsilon^2 \subset \IR^3$, around the singular point and one finds:
\begin{equation}
 \int_{S_\epsilon^2} ~  \Theta^{(I)}   ~=~   
 \begin{cases}  
 -4\,\pi\,  k_j^+   & \mbox{\quad for BPS} \\
 - 4\,\pi\, \big(h + \frac{q}{a_j} \big) \, k_j^- & \mbox{\quad for almost-BPS } 
  \end{cases}   
\label{Thetaint}
\end{equation}
Hence, the quantized dipole charges are
\begin{equation}
\widehat k_j ~\equiv~   \Big(h + \frac{q}{a_j} \Big) \, k^-_j
\label{effectivek}
\end{equation}
in the almost-BPS solutions.

The electrostatic potentials are relatively simple.  For BPS supertubes it is straightforward to apply the results of  \cite{Bena:2005va, Berglund:2005vb}:
\begin{eqnarray}
Z_1 &=& L_1 ~+~ \frac{K^2_+ \, K^3_+}{V}   ~=~ 1 ~+~ \frac{Q_2^{(1)}}{r_2} ~+~ \frac{Q_3^{(1)}}{r_3} ~+~ \frac{k^+_2\,k^+_3 }{ (h + \frac{q}{r})\,r_2\, r_3} \,,  \label{BPSZ1} \\
Z_2 &=& L_2~+~ \frac{K^1_+ \, K^3_+}{V}   ~=~  1  ~+~ \frac{Q_1^{(2)}}{r_1} ~+~ \frac{Q_3^{(2)}}{r_3} ~+~ \frac{k^+_1\,k^+_3 }{ (h + \frac{q}{r})\,r_1\, r_3} \,,  \label{BPSZ2} \\
Z_3 &=& L_3~+~ \frac{K^1_+ \, K^2_+}{V}   ~=~  1   ~+~ \frac{Q_1^{(3)}}{r_1} ~+~ \frac{Q_2^{(3)}}{r_2} ~+~ \frac{k^+_1\,k^+_2 }{ (h + \frac{q}{r})\,r_1\, r_2} \,. \label{BPSZ3}
\end{eqnarray}
For the almost-BPS supertubes one can use the results in \cite{Bena:2009en}:
\begin{eqnarray}
Z_1 &=& 1 ~+~ \frac{Q_2^{(1)}}{r_2} ~+~ \frac{Q_3^{(1)}}{r_3} ~+~ \Big(h + \frac{q \, r}{a_2 \, a_3} \Big)\,  \frac{k^-_2\,k^-_3 }{ r_2\, r_3} \,,  \label{nonBPSZ1} \\
Z_2 &=&  1  ~+~ \frac{Q_1^{(2)}}{r_1} ~+~ \frac{Q_3^{(2)}}{r_3} ~+~ \Big(h + \frac{q \, r}{a_1 \, a_3} \Big)\, \frac{k^-_1\,k^-_3 }{ r_1\, r_3} \,,  \label{nonBPSZ2} \\
Z_3 &=&  1   ~+~ \frac{Q_1^{(3)}}{r_1} ~+~ \frac{Q_2^{(3)}}{r_2} ~+~ \Big(h + \frac{q \, r}{a_1 \, a_2} \Big)\, \frac{k^-_1\,k^-_2 }{ r_1\, r_2} \,. \label{nonBPSZ3}
\end{eqnarray}
Note that the harmonic pieces, $L_I$,  are the same for both the BPS and almost-BPS systems.

The angular momentum vector, $k$, is decomposed in the usual manner:
\begin{equation}
k ~=~ \mu\, ( d\psi + A   ) ~+~ \omega \,.
\label{kansatz}
\end{equation}
The details will not be important here but for completeness we give them in Appendix A.

\subsection{The bubble equations}

One defines the usual symplectic inner products for the BPS and almost-BPS charges:
\begin{align}
&\Gamma_{12}^+  ~=~  k^+_1\, Q_2^{(1)}  - k^+_2\, Q_1^{(2)}   \,, \qquad  \Gamma_{13}^+  ~=~  k^+_1\, Q_3^{(1)}  - k^+_3\, Q_1^{(3)}   \,, \qquad\Gamma_{23}^+  ~=~  k^+_2\, Q_3^{(2)}  - k^+_3\, Q_2^{(3)}   \,,  \label{Gammadefns1} \\ 
&\widehat{\Gamma}_{12}  ~=~  \widehat k_1\, Q_2^{(1)}  - \widehat k_2\, Q_1^{(2)}   \,, \qquad  \widehat{\Gamma}_{13}  ~=~  \widehat k_1\, Q_3^{(1)}  - \widehat k_3\, Q_1^{(3)}   \,, \qquad \widehat{\Gamma}_{23}  ~=~  \widehat k_2\, Q_3^{(2)}  - \widehat k_3\, Q_2^{(3)}  \,.  \label{Gammadefns2}
\end{align}
Note that we have used the effective, local charges, $\widehat k_j$ in (\ref{Gammadefns2}).

It was shown in  \cite{Vasilakis:2011ki} that supertube regularity requires the locations of the supertubes to be related to the charges via:
\begin{eqnarray}
\frac{\Gamma^+_{12}}{|a_1 - a_2|} ~+~ \frac{\Gamma^+_{13}}{|a_1 - a_3|} &=& \frac{Q_1^{(2)} \, Q_1^{(3)}  }{k^+_1}  \Big(h + \frac{q}{a_1} \Big) ~-~   k^+_1 \,,  \label{BPSbubble1} \\
\frac{\Gamma^+_{21}}{|a_1 - a_2|} ~+~ \frac{\Gamma^+_{23}}{|a_2 - a_3|} &=&  \frac{Q_2^{(1)} \, Q_2^{(3)}  }{k^+_2}  \Big(h + \frac{q}{a_2} \Big) ~-~  k^+_2 \,,   \label{BPSbubble2} \\
\frac{\Gamma^+_{31}}{|a_1 - a_3|} ~+~ \frac{\Gamma^+_{32}}{|a_2 - a_3|} &=&   \frac{Q_3^{(1)} \, Q_3^{(2)}  }{k^+_3}  \Big(h + \frac{q}{a_3} \Big) ~-~  k^+_3 \,. \label{BPSbubble3}
\end{eqnarray}
for the BPS solutions.  For the almost-BPS supertubes one must impose
\begin{eqnarray}
\frac{\widehat\Gamma_{12}}{|a_1 - a_2|} ~+~ \frac{\widehat\Gamma_{13}}{|a_1 - a_3|} &=& \frac{Q_1^{(2)} \, Q_1^{(3)}  }{\widehat k_1}  \Big(h + \frac{q}{a_1} \Big) ~-~ \widehat k_1 ~-~ \epsilon_{123} \,  Y \,,  \label{nonBPSbubble1} \\
\frac{\widehat\Gamma_{21}}{|a_1 - a_2|} ~+~ \frac{\widehat\Gamma_{23}}{|a_2 - a_3|} &=&   \frac{Q_2^{(1)} \, Q_2^{(3)}  }{\widehat k_2}  \Big(h + \frac{q}{a_2} \Big) ~-~  \widehat k_2 ~-~ \epsilon_{213}   \,  Y \,,   \label{nonBPSbubble2} \\
\frac{\widehat\Gamma_{31}}{|a_1 - a_3|} ~+~ \frac{\widehat\Gamma_{32}}{|a_2 - a_3|} &=&   \frac{Q_3^{(1)} \, Q_3^{(2)}  }{\widehat k_3}  \Big(h + \frac{q}{a_3} \Big) ~-~ \widehat k_3 ~-~ \epsilon_{312}   \,  Y\,, \label{nonBPSbubble3}
\end{eqnarray}
where 
\begin{equation}
Y  ~\equiv~    \frac{h\,q \, k^-_1 k^-_2 k^-_3 }{a_1 a_2 a_3} ~=~    \frac{h\,q \, \widehat k_1 \widehat k_2 \widehat  k_3 }{(q+ h a_1) (q+ h a_2)  (q+ h a_3) } \,. \label{Ydefn}
\end{equation}
and
\begin{equation}
\epsilon_{ijk} ~\equiv~   \frac{(a_i - a_j)\,  (a_i - a_k)}{|a_i - a_j|\,  |a_i - a_k|}  \,. \label{epsdefn}
\end{equation}
The only difference between the two sets of equations is the term, $Y$, which vanishes in the supersymmetric limits of the solution ($h\rightarrow 0$ or $q \rightarrow 0$), and hence can be thought of as measuring the amount of supersymmetry breaking.

Our goal, in Section \ref{Bprobes}, will be to see how these bubble equations, and particularly the supersymmetry breaking term, can be obtained from a brane-probe calculation. 

\subsection{The Minkowski-space limit}
\label{MinkLim}

If one sets $h=0$ then the metric (\ref{TNmet}) becomes that of flat Minkowski space modded out by $\ZZ_q$.   In this limit, the BPS and almost-BPS solutions must become the same because both solutions are necessarily BPS since the base is flat.

It is evident that  the potential functions, (\ref{BPSZ1})--(\ref{BPSZ3})  and (\ref{nonBPSZ1})--(\ref{nonBPSZ3}), are identical for $h=0$.    The metric may then be written as 
\begin{equation}
ds^2_4 ~=~ q\, \Big[  \frac{dr^2}{r} ~+~ r \, \Big(   \Big(\frac{d\psi }{q}\Big)^2  ~+~  d\theta^2   ~+~ d\phi^2    ~\pm~ \frac{2 }{q} \cos \theta \,  d\psi \, d\phi) \Big)  \Big]    \,,
\label{flatmet}
\end{equation}
which takes a more canonical form if one sets $r =\frac{1}{4} \rho^2$.  The $\pm$ sign reflects the BPS, or almost-BPS description (\ref{Aform}) but this is a coordinate artifact.

There are several obvious ways of mapping the BPS description of the metric onto the almost-BPS description but one can most easily see the correct choice by looking at the magnetic fields, $B_\pm^{(I)}$,  for $h=0$.  Indeed, from (\ref{BIres}) one sees that the  substitution
\begin{equation}
\psi ~\rightarrow~ - q\,  \phi    \,, \qquad     \phi ~\rightarrow~ \frac{1}{q} \,  \psi   \,, \qquad  k_I^+ ~\rightarrow~ \widehat k_I
\label{BPStoNBPS}
\end{equation}
maps $ B^{(I)}_+   \rightarrow  B^{(I)}_-$.  This transformation also interchanges the choice of signs in (\ref{flatmet}).

In Appendix A we also verify that the transformation (\ref{BPStoNBPS}), for $h=0$, also maps the angular momentum vector in the BPS description to  that of almost-BPS description.

\section{Brane probes in almost-BPS solutions}
 \label{Bprobes}

\subsection{Brane probes}

We only require a relatively simple result from the brane probe analysis of \cite{Bena:2008dw} or the more recent work of \cite{Bena:2011fc}.  We need the so-called ``radius relation'' that determines the equilibrium position of a probe in a supergravity background.

One can use the M-theory frame or one can go back to  the approach of \cite{Bena:2008dw} and work with three-charge solutions in the IIA frame in which  the three electric charges, $N_1, N_2$ and $N_3$, of the solution  correspond to\footnote{For later convenience, we have permuted the labels relative to those of \cite{Bena:2008dw}.}:
\begin{equation}
N_1: \, {\rm D0}   \qquad  N_2: \,  {\rm F1} \, (z)  \qquad N_3: \, {\rm D4} \, (5678) \\,
\label{D0D4F1}
\end{equation}
where the numbers in the parentheses refer to spatial directions wrapped by the branes and $z\equiv x^{10}$. The  magnetic dipole moments of the solutions correspond to:
\begin{equation}
n_1: \, {\rm D6} \, (y5678z) \qquad  n_2: \,{\rm NS5} \, (y5678) \qquad n_3: \,
\,{\rm D2} \, (yz)  \,, \label{d6d2ns5}
\end{equation}
where $y$ denotes the brane profile in the spatial base, $(x^1, \dots,
x^4)$.  We will use a D2-brane probe, carrying electric D0 and F1
charges, $\widehat Q^{(1)}$ and $\widehat Q^{(2)}$, and with a
D2-dipole moment, $d_3$. For the supertube worldsheet coordinates, 
$\zeta^{\mu}$, we use static gauge and allow the supertube to wind
along both the $\psi$ and $\phi$ angles of the base space according
to:
\begin{equation} \label{staticgauge}
x^{0}=\zeta^0 \, , \, z=\zeta^1 \, , \, \psi=\nu_\psi\zeta^2 \, , \, \phi=\nu_\phi\zeta^2 \,.
\end{equation}
For such a probe, sitting along the positive $z$-axis of the base metric, the radius relation is given by:
\begin{equation}
\big[\widehat Q^{(1)}  +  d_3  \, \cB^{(2)} \big] \,\big[\widehat Q^{(2)}  +  d_3  \, \cB^{(1)} \big]  ~=~ d_3^2\,  \frac{Z_3}{V}\left(\nu_\psi + q\nu_\phi \right)^2 \,, \label{radreln}
\end{equation}
where $\cB^{(I)}$ is the pullback of the vector field, $B^{(I)}$, onto the  profile of the supertube on the spatial base:
\begin{equation}
\cB^{(I)} ~=~ B^{(I)}_\mu \, \frac{\partial x^\mu}{\partial \zeta_2} \,.  \label{pullback}
\end{equation}
%
Then, using (\ref{BIres}), for the BPS solutions:
\begin{equation}
\cB^{(I)} ~=~ \cB^{(I)}_+  ~=~   \frac{k_I^+}{(h + \frac{q}{r})\,r\, r_I} \, \big[\,  \nu_\psi \, r   +   q\,  a_I \,    \nu_\phi  -    h\,r\, (r \cos \theta - \, a_I )\, \nu_\phi \,  \big]    \,, \label{cBBPS}
\end{equation}
while for non-BPS we have
\begin{equation}
\cB^{(I)} ~=~ \cB^{(I)}_-  ~=~    \frac{k_I^- }{a_I \, r_I}\, \big[\, a_I \,\nu_\psi    -   q \,  r  \,  \nu_\phi -  h\,a_I \,   (r \cos \theta  - a_I)\, \nu_\phi \,  \big]     \,. \label{cBNBPS}
\end{equation}

In deriving this radius relation we have set the worldvolume field
strength $\cF_{tz} = +1$, which is the choice one makes when placing
supersymmetric probes in supersymmetric solutions, and one may ask
whether this choice is justified for a non-BPS probe supertube,
especially because this choice does not describe the metastable
supertube minima of \cite{Bena:2011fc,Bena:2012zi}. There are two ways
to see that this choice  correctly reproduces the almost-BPS probe
supertube minima. The first, and most direct, would be to evaluate the Hamiltonian derived in
\cite{Bena:2011fc} in an almost-BPS solution, and to find directly
that the minima of this Hamiltonian have $\cF_{tz} = +1$.  While he have not done
the former computation, there is a simpler,  second argument that shows 
what the outcome must be.  Remember that, unlike non-extremal solutions, almost-BPS
solutions have the mass and the charge equal, and the only way a
supertube action can yield a minimum with the mass equal to the charge
is if its worldvolume field strength satisfies $\cF_{tz} = +1$.

\subsection{Probing a supergravity solution}
\label{threetubes}

We now replace the third supertube in the solution of Section \ref{Sect:Sols} by a probe: we set $k_3^\pm = Q_3^{(1)} = Q_3^{(2)} =0$ and consider the action of a probe with charges $\widehat Q_{probe}^{(1)}$, $\widehat Q_{probe}^{(2)}$  and dipole charge  $d_3$ at a location, $a_3$, on the $z$-axis and winding once around the $\psi$-fiber.

We then have:
\begin{equation}
\cB^{(I)}_+  ~=~   \frac{k_I^+}{(h + \frac{q}{a_3}) \, |a_I - a_3|}  \,,  \qquad \cB^{(I)}_-  ~=~    \frac{k_I^- }{ |a_I - a_3|}     \,.   \label{cBres} 
\end{equation}
We know from the results of \cite{Bena:2008dw} that for the BPS choice the radius relation, (\ref{radreln}), indeed yields the bubble equation (\ref{BPSbubble3}).  We therefore focus on the almost-BPS solution.

Inserting  $\cB^{(I)} =  \cB^{(I)}_-$ in (\ref{radreln}) and using (\ref{cBres}) one can rearrange the radius relation to give
\begin{align}
\frac{1}{|a_1 -a_3|} \Big[ d_3 \, Q^{(3)}_1 ~-~  &   \Big(h+ \frac{q}{a_3} \Big)\, k^- _1 \,\widehat Q^{(1)}_{probe}  \Big] ~+~ \frac{1}{|a_2 -a_3|} \Big[ d_3 \, Q^{(3)}_2  ~-~   \Big(h+ \frac{q}{a_3} \Big)\, k^- _2 \,\widehat Q^{(2)}_{probe}   \Big] \\
&  ~=~ \Big(h+ \frac{q}{a_3} \Big)\,   \frac{\widehat Q^{(1)}_{probe}  \, \widehat Q^{(2)}_{probe} }{d_3} ~-~  d_3 ~+~  \frac{q\, k^- _1\, k^- _2\, d_3 }{|a_1 -a_3|\, |a_2 -a_3|} \, \Big( \frac{a_1 \, a_2-a_3^2 }{a_1 \, a_2 \, a_3} \Big)  \,.
 \label{Radreln1} 
\end{align}
We now explicitly use the ordering (\ref{ajorder}) to write $|a_i - a_j|  = a_i - a_j$ for $i> j$ and we use the definition (\ref{effectivek}) to obtain:
\begin{align}
\frac{1}{|a_1 -a_3|} \Big[ d_3 \, Q^{(3)}_1 ~-~  &  \widehat k_1 \, Q^{(1)}_{probe}  \Big] ~+~ \frac{1}{|a_2 -a_3|} \Big[ d_3 \, Q^{(3)}_2  ~-~  \widehat  k _2 \,  Q^{(2)}_{probe}   \Big] \nonumber \\
&  ~=~ \Big(h+ \frac{q}{a_3} \Big)\,   \frac{  Q^{(1)}_{probe}  \,  Q^{(2)}_{probe} }{d_3} ~-~  d_3 ~-~  \frac{h\, q\, k^- _1\, k^- _2\, d_3 }{a_1 a_2 a_3\,(h+ \frac{q}{a_3} )}    \,,
 \label{Radrelnres} 
\end{align}
where we have introduced 
\begin{equation}
 Q^{(1)}_{probe}   ~\equiv~   \widehat  Q^{(1)}_{probe}  ~-~   \frac{q\, \widehat{k}_2 \,  d_3 }{a_2 a_3\,(h+ \frac{q}{a_2} )(h+ \frac{q}{a_3} )}  \,, \qquad  Q^{(2)}_{probe}   ~\equiv~   \widehat  Q^{(2)}_{probe}  ~-~   \frac{q\, \widehat{k}_1 \,  d_3 }{a_1 a_3\,(h+ \frac{q}{a_1} )(h+ \frac{q}{a_3} )}  \,.
 \label{chgshift1} 
\end{equation}
This exactly matches the bubble equation (\ref{nonBPSbubble3}) provided one makes the identifications:
\begin{equation}
d_3 ~=~ \widehat k_3 ~=~   \Big(h+ \frac{q}{a_3} \Big)\, k_3^-    \,, \qquad    Q^{(I)}_{probe}  ~=~    Q^{(I)}_{3} \,.
 \label{idents} 
\end{equation}
Hence, the action of a probe supertube that is placed at the outermost
position of an almost-BPS solution that contains two other supertubes
of different species can capture exactly the supergravity information
about the location of this supertube. Since all the terms in the
supertube bubble equations only contain two- and three-supertube
interactions, we will see in  Section \ref{multitubes} that this
result can be straightforwardly generalized to a
probe supertube placed at the exterior position of an almost-BPS
solution containing an arbitrary number of supertubes of arbitrary
species.

There is another more compact way to write the formula that gives the shift from the quantized charges to the BPS parameters
\begin{equation}
  Q^{(1)}_{probe}   ~=~   \widehat  Q^{(1)}_{probe}  ~-~   \frac{q\, \widehat{k}_2 \,  \widehat{k}_3  }{a_2 \,  a_3 V_2 V_3}  \,, \qquad  Q^{(2)}_{probe}   ~=~   \widehat  Q^{(2)}_{probe}  ~-~   \frac{q\, \widehat{k}_1 \, \widehat{k}_3 }{a_1 \,  a_3 V_1 V_3}  \,.
 \label{chgshift2} 
\end{equation}

\subsection{Interpretation of the charge shift}
\label{interpretation}

Given that we have matched the probe calculation to the supergravity one by postulating the charge shift above, it is legitimate to ask whether this charge shift has any direct physical interpretation apart from the fact that it maps the probe result onto the supergravity result. We will argue in the remaining part of the paper that this charge shift encodes very non-trivial properties of almost-BPS solutions. 

We  begin by comparing almost-BPS solutions to BPS multi-center solutions (for which there is no charge shift \cite{Bena:2008dw}) in the limits when the two kinds of solutions become identical. We will then explain how this charge shift solves an old puzzle about the relation between quantized charges and moduli in almost-BPS solutions

\subsubsection{Mapping BPS to non-BPS solutions}

There are two ways to turn an almost-BPS solution into a BPS one. The first is to set the Taub-NUT charge, $q$,  to zero (and obtain black strings and supertubes extended along the $S^1$ of $\IR^3 \times S^1$), and the second is to set the constant, $h$, in the Taub-NUT harmonic function equal to zero. In the first limit ($q=0$), the charge shifts automatically vanish, consistent with the result of \cite{Bena:2008dw}.

However, for $h=0$, the shift does not vanish and this  may seem a little puzzling. To understand the origin of the shift we should remember that, in the $h=0$ limit, the BPS and the almost-BPS solutions are identical up to the coordinate transformation  (\ref{BPStoNBPS}). Hence, a supertube wrapping the $\psi$-fiber once in the almost-BPS solution and a supertube wrapping the $\psi$-fiber once in the BPS writing of the same solution wrap different circles. Indeed, the supertube that wraps the $\psi$-fiber once in the almost-BPS sozflution has $(\nu_\psi,\nu_\phi) = (1,0)$.  Then the transformation (\ref{BPStoNBPS}) means that in the BPS writing of the solution this object is a supertube that wraps the $\phi$-fiber $\frac{1}{q}$ times, and hence it has $(\nu_\psi,\nu_\phi) = (0,\frac{1}{q})$. 

Clearly 
two supertubes with different windings are not the same object, and the only way their radius relations can be the same is if their charges are different. To be more precise, when $h=0$,  (\ref{cBBPS}) reduces to:
\begin{equation}
\cB^{(I)}_+  (\nu_\psi,\nu_\phi) ~=~   \frac{k_I^+}{q \,(a_3 - a_I)} \, [  \nu_\psi \, a_3   +    \nu_\phi   \, q\,  a_I   ]~=~   \frac{k_I^+\,  (\nu_\psi + q\, \nu_\phi  ) \, a_3}{q \,(a_3 - a_I)}  ~ - ~  k_I^+\, \nu_\phi      \,, \label{cBBPSsimp}
\end{equation}
where we have used (\ref{ajorder}).  The first term is identical for either choice of winding numbers:  $(\nu_\psi,\nu_\phi) = (1,0)$ and $(\nu_\psi,\nu_\phi) = (0,\frac{1}{q})$, and this makes sense because, for either choice, the Taub-NUT fiber is being wrapped once.   However, the last term is different, which indicates that the almost-BPS supertube also wraps the Dirac string of the background magnetic flux. Hence, if the supertubes with $(\nu_\psi,\nu_\phi) = (1,0)$ and $(\nu_\psi,\nu_\phi) = (0,\frac{1}{q})$ are to have the same radius 
their charges must be related by a shift. Using the charge identification $k_I^+ \rightarrow \widehat k_I $ in (\ref{BPStoNBPS})  and using (\ref{effectivek}) for $h=0$, one sees that this charge shift is simply:
\begin{equation}
\widehat Q^{(1)}_{(1,0)} ~ = ~ \widehat Q^{(1)}_{(0,\frac{1}{q})} ~ - ~  \frac{d_3 \, \widehat{k}_2}{q} \,, \qquad  \widehat Q^{(2)}_{(1,0)} ~ = ~ \widehat Q^{(2)}_{(0,\frac{1}{q})} ~ - ~    \frac{d_3 \, \widehat{k}_1}{q}  \,. \label{DSchgshift1}
\end{equation}
which exactly matches the shift in  (\ref{chgshift1}) for $h=0$.  Hence the shift between the almost-BPS supergravity charge and the almost-BPS quantized charge is the same as the shift from the BPS quantized charge to the almost-BPS quantized charge. This implies that the BPS quantized and supergravity charges are the same as the almost-BPS supergravity charge. 

Another way to understand this result is to remember that one can consider a solution with a probe supertube of a certain charge wrapping a Dirac string, and take away the Dirac string by a large gauge transformation. As explained in \cite{Bena:2008dw}, this changes the charges of the supertube. If one does this in our situation, and takes the Dirac string away from the location of the probe supertube, both the BPS and the almost-BPS supertubes will have the same charge and wrapping and will become the same object. Clearly, upon back-reaction the supergravity charges of the two will be the same. Hence, the reason why the quantized charge of the almost-BPS supertube is shifted from its supergravity charge is because, when written as a probe in $\IR^4$, this supertube wraps  a Dirac string non-trivially. 

Note that when $h=0$ this shift involves only integers, but for a generic almost-BPS solution this shift depends also on the moduli. As we will explain in section \ref{multitubes2} this happens because almost-BPS solutions have an additional magnetic flux on the non-compact cycle extending to infinity.

\subsubsection{A puzzle about charges and its resolution}

Once one understands how the quantized charges of the centers of an almost-BPS solution are related to the supergravity charge parameters one can re-examine the problem of moduli-dependence of the asymptotic charge of almost-BPS solutions. Recall  that for a BPS solution in Taub-NUT the warp factors go asymptotically like
\begin{eqnarray}
  &Z^{(1)}   ~\sim~   \Big[ Q_2^{(1)} + Q_3^{(1)} \Big] \,  \displaystyle{\frac{1}{r}}     \,, \qquad Z^{(2)} ~\sim~      \Big[ Q_1^{(2)} + Q_3^{(2)} \Big] \, \frac{1}{r}   &\,,\nonumber \\
  &Z^{(3)}   ~\sim~   \Big[ Q_1^{(3)} + Q_2^{(3)} \Big] \, \displaystyle{\frac{1}{r}}   \,, &
 \label{BPSasymp} 
\end{eqnarray}
which confirms the fact that the asymptotic charges are simply the sum of the quantized charges of the individual component supertubes. However, for an almost-BPS solution one has
\begin{eqnarray}
  &Z^{(1)}   ~\sim~     \Big[ Q_2^{(1)} + Q_3^{(1)} +  \frac{q\, \widehat{k}_2 \,  \widehat{k}_3  }{a_2 \,  a_3 V_2 V_3} \Big] \, \displaystyle{\frac{1}{r}}     \,, \qquad Z^{(2)} ~\sim~      \Big[ Q_1^{(2)} + Q_3^{(2)} +  \frac{q\, \widehat{k}_1 \,  \widehat{k}_3  }{a_1\,  a_3 V_1 V_3} \Big] \,  \displaystyle{\frac{1}{r}} &  \,,\nonumber \\ 
  &Z^{(3)}   ~\sim~     \Big[ Q_1^{(3)} + Q_2^{(3)} +  \frac{q\, \widehat{k}_1 \,  \widehat{k}_2  }{a_1 \,  a_2 V_1 V_2} \Big] \, \displaystyle{\frac{1}{r}}  \,. 
 \label{NBPSasymp} 
\end{eqnarray}
This appears to imply that if one changes the moduli of a solution continuously, while keeping the charges of the centers fixed, the asymptotic charges will change continuously. If this were true, an almost-BPS solution would have non-quantized asymptotic charges for generic values of the moduli, and would only exist on certain submanifolds of the moduli space where the asymptotic charges are integer. Needless to say, this would be rather peculiar.

Our results show that this does not, in fact, happen. If one expresses the supergravity charge in terms of quantized charges, the moduli-dependent term drops out and the asymptotics become identical to the one in (\ref{BPSasymp}). Hence, the asymptotic charges of almost-BPS solutions are also equal to the sum of the quantized charges of the centers, and the quantization of charges does not restrict the the moduli space of supertube locations.

To see this we can make the following Gedankenexperiment: we start with a single supertube of species $I=1$ at position $a_1$, for which the supergravity charges are the same as the quantized charges
\begin{equation}
Q_1^{(2)}=\widehat{Q}_1^{(2)}\,, \qquad Q_1^{(3)}=\widehat{Q}_1^{(3)}\,.
\end{equation}
We then bring in a second supertube of species $I=2$ at position $a_2$. The supergravity charge $Q_2^{(3)}$ of this supertubes is shifted from its quantized charge $\widehat{Q}_2^{(3)}$ because of the presence of the first supertube, while its other charge remains the same:
\begin{equation}
Q_2^{(1)}=\widehat{Q}_2^{(1)}\,, \qquad Q_2^{(3)}=\widehat{Q}_2^{(3)}-\frac{q \widehat{k}_1^{(1)} \widehat{k}_2^{(2)}}{a_1 a_2 V_1 V_2}\,,
\end{equation}

 The asymptotic charges $\widehat{\mathcal{Q}}^{(I)}_{12}$ of the resulting solution are given by the (sum of the) quantized charges, $\widehat{\mathcal{Q}}_{12}^{(1)}=\widehat{Q}_2^{(1)}=Q_2^{(1)}$,  $\widehat{\mathcal{Q}}_{12}^{(2)}=\widehat{Q}_1^{(2)}=Q_1^{(2)}$ and
%
\begin{equation}
\widehat{\mathcal{Q}}_{12}^{(3)}=\widehat{Q}_1^{(3)}+\widehat{Q}_2^{(3)}={Q}_1^{(3)}+{Q}_2^{(3)}+\frac{q \widehat{k}_1^{(1)} \widehat{k}_2^{(2)}}{a_1 a_2 V_1 V_2}\,,
\end{equation}
which agrees with the coefficient of $r^{-1}$  in $Z^{(3)}$ in (\ref{NBPSasymp}).
After back-reacting the second supertube we bring in a third supertube of species $I=3$ at position $a_3$ whose supergravity charges after back-reaction are shifted  with respect to their quantized charges
\begin{equation}
Q_3^{(1)}=\widehat{Q}_3^{(1)}-\frac{q \widehat{k}_2^{(2)}\widehat{k}_3^{(3)}}{a_2 a_3 V_2 V_3}\,, \qquad Q_3^{(2)}=\widehat{Q}_3^{(2)}-\frac{q  \widehat{k}_1^{(1)}\widehat{k}_3^{(3)}}{a_1 a_2 V_1 V_2}\,,
\end{equation}
because of the other supertubes. The asymptotic charges $\widehat{\mathcal{Q}}^{(I)}_{123}$ of the solution with all three supertubes back-reacted are $\widehat{\mathcal{Q}}_{123}^{(3)}=\widehat{\mathcal{Q}}_{12}^{(3)}\sim r ~Z^{(3)}$ and
\begin{eqnarray}
 &&\widehat{\mathcal{Q}}_{123}^{(1)}=\widehat{\mathcal{Q}}_{12}^{(1)}+\widehat{Q}_3^{(1)}={Q}_2^{(1)}+{Q}_3^{(1)}+\frac{q \widehat{k}_2^{(2)} \widehat{k}_3^{(3)}}{a_2 a_3 V_2 V_3}\sim r ~Z^{(1)}\,,\\
  &&\widehat{\mathcal{Q}}_{123}^{(2)}=\widehat{\mathcal{Q}}_{12}^{(2)}+\widehat{Q}_3^{(2)}={Q}_1^{(2)}+{Q}_3^{(2)}+\frac{q \widehat{k}_1^{(1)} \widehat{k}_3^{(3)}}{a_1 a_3 V_1 V_3}\sim r ~Z^{(2)}\,.
\end{eqnarray}
As advertised, these charges match the coefficients of $r^{-1}$
in the $Z^{(I)}$. Hence, while the shift between the quantized and
supergravity charges might have seemed surprising at first, it represents 
 the missing ingredient necessary to relate the asymptotic
charge of a multi-center solution to those of the centers.

One can also use these equations to determine the shift between the
quantized and supergravity charges for an almost-BPS black ring in
Taub-NUT \cite{Bena:2009ev}. Although black rings have no DBI
description, one can make a black ring by bringing together three
supertubes with three different types of dipole charges. Since in this
process both the supergravity and the quantized charges are preserved,
the shifts of the black ring charges will be the shifts of
the charges of the composing supertubes. For a Taub-NUT almost-BPS
black ring with dipole charges $\widehat{k}^{(I)}$ at position $a_1$
these shifts are:
\begin{equation}
Q^{(I)}= \widehat{Q}^{(I)} - {|\epsilon_{IJK}|} \frac{q \widehat{k}^{(J)} \widehat{k}^{(K)}}{a_1^2  V_1^2}\, ,
\end{equation}
where capital Latin indices are not summed.
Note that there are $n!$ ways of making a black ring starting with $n$
supertubes, corresponding to the different relative orderings of these
supertubes, and for each ordering the charge shifts of various
supertubes are different, but the sum of all these charge shifts (which
gives the black ring charge shift) is always the same.

We are now ready to further use this result and the iterative
procedure outlined above in order to unambiguously determine the
relation of the supergravity charge parameters and the quantized
charges in a solution with an arbitrary number of centers.

\subsection{Generalization to many supertubes}
 \label{multitubes}

We can generalize the calculation in the previous section to an almost-BPS solution containing an arbitrary number of colinear supertubes. Without loss of generality we consider a solution containing $i-1$ supertubes at positions
\begin{equation} \label{norder1}
0<a_1<a_2<.....<a_{i-1} \,,
\end{equation}
and we bring a probe supertube of species $I$ (wrapping the tori $T_J^2$ and $T_K^2$ in (\ref{11Dmetric})) to the point $a_i$ on the z-axis, at the outermost position with respect to the other supertubes ($a_i > a_{i-1}$). This probe has electric charges $\widehat Q_i^{(J)}$, $\widehat Q_i^{(K)}$ and dipole charge $d_i^{(I)}$. Starting from the radius relation:
\begin{equation} \label{RRmulti}
\left[\widehat Q_i^{(J)}+d_i^{(I)} \mathcal{B}_i^{(K)}\right]\left[\widehat Q_i^{(K)}+d_i^{(I)}\mathcal{B}_i^{(J)}\right]=(d_i^{(I)})^2\frac{Z^{(I)}_i}{V_i}
\end{equation}
with
\begin{equation} \label{Bminusmulti}
\mathcal{B}^{(L)}_{i}=\mathcal{B}^{(L)}_-(r=a_i)=\sum_l \frac{k_l^{-(L)}}{|a_l-a_i|}=\sum_l \frac{\widehat{k}_l^{(L)}}{|a_l-a_i|\, V_l}\, \quad ,\qquad \quad V_i = h+\frac{q}{a_i}\,,
\end{equation}
and  
\begin{equation} \label{ZNBPSmulti}
Z^{(I)}_i = Z^{(I)}(r=a_i) = 1 + \sum_j \frac{Q_j^{(I)}}{|a_i-a_j|} + |\epsilon_{IJK}|\sum_{j,k}\left( h+\frac{qa_i}{a_ja_k}\right)\frac{k_j^{-(J)} k_k^{-(K)}}{|a_i-a_j||a_i-a_k|}\,, 
\end{equation}
we obtain the shift of charges by matching this relation to the bubble equation for the back-reacted probe center (that can be derived from the results of \cite{Bena:2009en}):
\begin{eqnarray} \label{multibubble}
&&\hspace{-3cm}\Big\{\sum_{j} \frac{1}{|a_i-a_j|} \big[ d_i^{(I)} Q_j^{(I)} - \widehat{k}_j^{(J)}  Q_i^{(J)}  \big]  +
\sum_{k} \frac{1}{|a_i-a_k|} \big[ d_i^{(I)} Q_k^{(I)} - \widehat{k}_k^{(K)}  Q_i^{(K)}  \big]
\Big\}\nonumber \\  &=& \frac{Q_i^{(J)} Q_i^{(K)}}{d_i^{(I)}} \left( h+\frac{q}{a_i} \right) - d_i^{(I)} - 
|\epsilon_{IJK}|
\frac{hqd_i^{(I)}}{a_i\left(h+\frac{q}{a_i}\right)} \sum_{j,k} \epsilon_{ijk} \frac{k_j^{-(J)} k_k^{-(K)}}{a_j a_k}\,,
\end{eqnarray}
where there is no summation over capital Latin indices\footnote{Note that the terms involving $|\epsilon_{IJK}|$ in $Z^{(I)}$ and the bubble equation \eqref{multibubble} differ by a factor $1/2$ from the usual form (see, for example, \cite{Bena:2009en}) where {\it there is} summation over $I,J,K$.}.
We find that, upon identifying $d_i^{(I)}=\widehat{k}_i^{(I)}= k_i^{-(I)}V_i$, the shift between the supergravity charge parameters of the supertube and its quantized charges are
\begin{equation} \label{multishiftonsign}
Q_i^{(J)} = \widehat Q_i^{(J)} - {|\epsilon_{JKI}|} \sum_{k} \frac{q\widehat{k}_k^{(K)}\widehat{k}_i^{(I)}}{a_k a_i V_k V_i} \, 
\end{equation}
and similarly for $Q_i^{(K)} $, where there is no summation over capital Latin indices.  In equations (\ref{multibubble})--(\ref{multishiftonsign}) the indices $j,k$ run over the positions of supertubes of species $J,K$ respectively.

One can now use this formula and find the charge shifts for all the centers of a certain solution, by constructing it recursively by bringing in probe supertubes. From the previous analysis it is clear that the relation between the supergravity and the quantized charges of a given center in the interior of the solution does not change as one brings more and more supertubes at the outermost positions. Hence, the shifts of the charges of a given supertube center only depend on the locations and dipole charges of the centers that are between this supertube center and the Taub-NUT center, but not on the locations or charges of the centers at its exterior:
\begin{equation} \label{multishiftonsign2}
Q_i^{(J)} = \widehat Q_i^{(J)} -{|\epsilon_{JKI}|} \sum_{k<i} \frac{q\widehat{k}_k^{(K)}\widehat{k}_i^{(I)}}{a_k a_i V_k V_i} .
\end{equation}

It is not hard to see that the asymptotic charge of the solution (which can be read off from the asymptotics of $Z^{(I)}$) is now the sum of the quantized charges of all the centers
\begin{equation}
\widehat{\mathcal{Q}}^{(I)}_{\infty}
= \sum_{j} Q_j^{(I)}+|\epsilon_{IJK}|\sum_{j,k}\frac{q\widehat{k}_j^{(J)} \widehat{k}_k^{(K)}}{a_ja_kV_j V_k}
= \sum_i\widehat{Q}_i^{(I)} .
\end{equation}

One can also extend this calculation to describe a configuration containing an arbitrary collection of concentric supertubes and black rings, by constructing the black rings from three different species of supertubes. The general formula relating the supergravity charge parameters to the quantized charges is
\begin{equation} \label{multi-general}
Q_i^{(J)} = \widehat Q_i^{(J)} -{|\epsilon_{JKI}|} \sum_{k<i} \frac{q\widehat{k}_k^{(K)}\widehat{k}_i^{(I)}}{a_k a_i V_k V_i} 
 - {|\epsilon_{JKI}|} \frac{q \widehat{k}_i^{(K)} \widehat{k}_i^{(I)}}{a_i^2  V_i^2}\,,
\end{equation}
where once again the capital latin indices are not being summed.

\section{Extracting the complete supergravity data from supertubes}
 \label{SGdata}

We have seen that if one considers an axially-symmetric brane configuration and brings in a brane probe along the axis from one side of the configuration then the radius relation of that probe exactly reproduces  the bubble equation for the charge center that would replace the brane probe in a fully  back-reacted  supergravity configuration.   This correspondence requires the charge shifts described in Section  \ref{Bprobes}.  Since we are considering only the action of the probe, there is no immediate way in which this action could directly yield the fully back-reacted bubble equations for the other centers in response to the probe.   That is, given the  $n$-supertube solution, the probe radius relation for the $(n+1)^{\rm st}$ supertube yields the exact supergravity bubble equation for that supertube.  However, when back-reacted, the probe supertube must introduce modifications to the bubble equations for the other supertubes and these are not given directly by the computations described above.

In contrast,  the analysis of brane probes in BPS solutions \cite{Bena:2008dw} required no charge shifts and the bubble equations contained only two-body interactions  of the form ${\Gamma_{ij}}{|\vec y_i - \vec y_j |}^{-1}$  as in   (\ref{Gammadefns2})--(\ref{BPSbubble3}).  In that solution, the action of the probe was used to read off all the two-by-two terms between the probe and the centers, and thus derive the bubble equations for {\it all} the centers. However, for almost-BPS solutions the bubble equations for a certain center contain complicated three-body terms, which cannot be read off from the action describing the location of another center. However, there is still a (more complicated) way to recover the full supergravity data from DBI actions: given a solution with $(n+1)$ supertubes one can examine all the ways of extracting one supertube and treating it is a probe in the background sourced by the others. As we will show below, this yields all the bubble equations of the solution. 

\subsection{Reconstructing the bubble equations from probes}
 \label{ProbesGiveBubbles}

For non-BPS supertubes  there are  three elements for the inductive modification of the bubble equations as one goes from $n$ to $(n+1)$ supertubes:
\begin{itemize}
\item[(i)]  Compute the charge shifts of all probe charges.
\item[(ii)]  Write  the left-hand sides of the bubble equations with all two-body interactions, ${\Gamma_{ij}}{|a_i - a_j |}^{-1}$.
\item[(iii)]  Compute and include all the supersymmetry breaking terms, $Y$, as in (\ref{Ydefn}) and include them with the correct relative sign on the  right-hand sides of the bubble equations as in (\ref{nonBPSbubble1})--(\ref{nonBPSbubble3}).
\end{itemize}

If one assembles the configuration by bringing in the charges successively along the symmetry axis from the right then the charge shifts are given by (\ref{multishiftonsign}).  The  two-body interactions are fixed iteratively exactly as in the BPS solution.  These terms are also fixed by the consistency condition  that the sum of all the bubble equations must be identically zero.  The new  feature is that we need to specify the algorithm for (iii):  We need to generalize the supersymmetry breaking terms and how they are to be included in the bubble equations.  It is clear that the bubble equation for a supertube of type $I$ will have an interaction, $Y$, of the form  (\ref{Ydefn}) with every pair of supertubes of species $J,K$ where $I,J$ and $K$ are all distinct.  The issue is to introduce these terms with the correct signs in each bubble equation.   
One can do this by making the more formal induction using all the ways of extracting one supertube and treating it is a probe.  However, for simplicity, we will start by restricting our attention to configurations in which the probe is always the outermost supertube.  As we will see in the next section, the charge shifts are modified when the probe is some other supertube in the configuration and so the discussion is a little more complicated.  We will begin by completing the    discussion  of Section  \ref{multitubes} by giving a recipe for the supersymmetry breaking terms when the probe is  the outermost supertube and we will explain its derivation below.

Suppose  we have an $n$-supertube system and we bring in, as a probe,  a supertube of species $I$ then the bubble equations for a center of type $J$ at position $a_j$  must include an extra supersymmetry breaking term, $Y^{(J,I)}_{j, (n+1)}$,  that is to be subtracted from the right-hand side. These terms are given by:
\begin{equation} \label{extraY}
\begin{split}
 Y^{(J,I)}_{j, (n+1)} & =|\epsilon_{IJK}| \sum_k \frac{\left(a_j-a_{n+1}\right)\left(a_j-a_k\right)}{|a_j-a_{n+1}||a_j-a_k|} \frac{hqd^{(I)}_{n+1}\hat{k}^{(J)}_j\hat{k}^{(K)}_k}{a_{n+1}a_ja_kV_{n+1}V_jV_k} = \\
&= - |\epsilon_{IJK}| \sum_k \frac{\left(a_j-a_k\right)}{|a_j-a_k|} \frac{hqd^{(I)}_{n+1}\hat{k}^{(J)}_j\hat{k}^{(K)}_k}{a_{n+1}a_ja_kV_{n+1}V_jV_k}
\end{split}
\end{equation}
where, for fixed $I$ and $J$, the summation over $k$ runs only over supertubes of species $K$, with $I,J, K$ all distinct. In the last equality we used that $a_{n+1}>a_j$ for $j=1,2,...,n$. Thus, as we assemble the system by successively bringing supertubes from the right, for every new supertube added we have to subtract (\ref{extraY}) from the right hand side of the bubble equations of the previously back-reacted system. 
 
\subsection{Assembling colinear supertubes in general} 
\label{colineargeneral}

The general bubble equations (\ref{multibubble}) for a system of $n$ colinear supertubes contain supersymmetry breaking terms, $Y$, that come with a relative sign given by by $\epsilon_{ijk}$ defined in (\ref{epsdefn})\footnote{This $\epsilon_{ijk}$ should not be confused with the factors of $|\epsilon_{IJK}|$  that give the triple intersection number of the dipole charges.}. In the procedure discussed in Section 3,  we always had $\epsilon_{ijk}=+1$ since the probe was always being placed in the outermost position of the back-reacted geometry.  However, if the probe is of species $I$ and located at a general position, $a_i$, then we find that its radius relation can be mapped directly onto is supergravity bubble equation if one uses the following more general relation between supergravity and quantized brane charges:
\begin{equation} \label{multishift}
Q_i^{(J)} = \widehat Q_i^{(J)} - |\epsilon_{JKI}|\sum_{k} \frac{a_i-a_k}{|a_i-a_k|}\frac{q\widehat{k}_k^{(K)}\widehat{k}_i^{(I)}}{a_k a_i V_k V_i} \,,
\end{equation}
where, once again, the capital latin indices are not being summed and the sum over $k$ runs over supertubes of species $K$. Thus for $a_i>a_k$ the shift terms get subtracted from the quantized charge as before, but for $a_i<a_k$ the shift terms have to be added! We also find that if one uses  the more general shifts in  (\ref{multishift})  then the radius relation also exactly reproduces  precisely the correct supersymmetry breaking terms, Y, in the bubble equations.  

\begin{figure}[t]
 \centering
    \includegraphics[width=10cm]{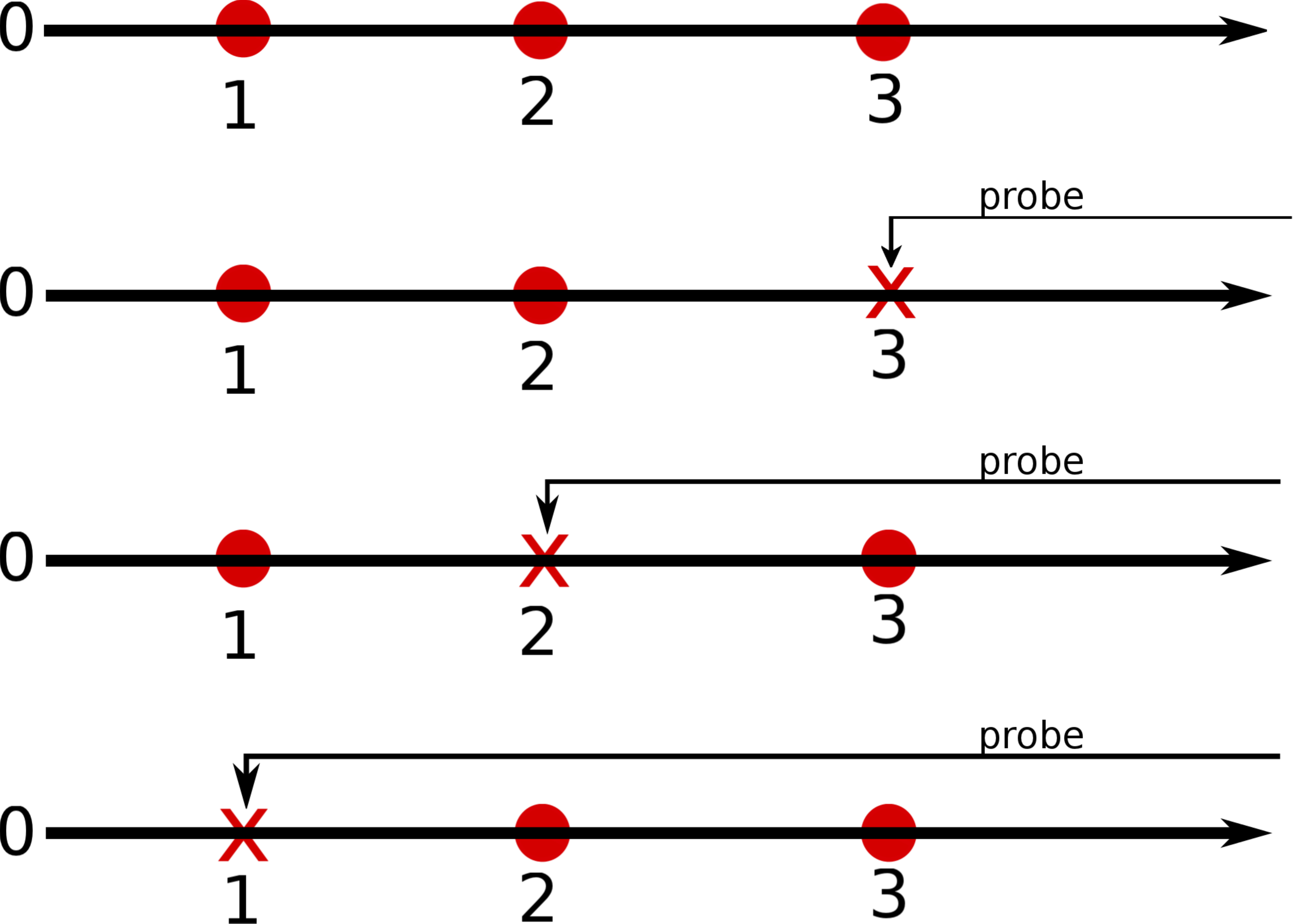}
    \caption{\it \small The first graph represents the three-supertube back-reacted geometry while the three below represent the iterative procedure by which one removes each center of the system and treats it as a probe in the background of the remaining two supergravity supertubes. Because supertubes cross over each other we need the formula (\ref{multishift}) to generate the shift with the correct sign.}
\label{iterative}
\end{figure}

As a result, by using  (\ref{multishift}), we can recreate the whole set of bubble equations of the system by the following iterative procedure. Consider a $(n+1)$-supertube system with preassigned order
\begin{equation} \label{norder}
0<a_1<a_2<.....<a_{n-1}<a_{n+1} \,.
\end{equation}
Now consider each supertube, in turn, as a probe in the supergravity
background of the remaining $n$ and imagine placing the probe in its
assigned position, (\ref{norder}), to recreate the full
$(n+1)$-supertube system.  To be more explicit, start by taking the
supertube at $a_{n+1}$ out of the supergravity system and treat it as
a probe being placed at position $a_{n+1}$. This generates the bubble
equation for the center at $a_{n+1}$ as described earlier. Next,
imagine the supergravity system where the supertube at position
$a_{n}$ has been removed and replaced by a probe placed at position
$a_{n}$.  The radius relation of this probe produces exactly the
bubble equation for the center at $a_{n}$ provided one uses the charge
shifts defined in (\ref{multishift}). One can then repeat this
iterative procedure (illustrated in Fig. \ref{iterative}) for each of
the $n+1$ centers of the system.

As we noted earlier, for a probe of species $I$, the bubble equations
for the center of type $J$ will involve a supersymmetry breaking term,
$Y^{(J,I)}_{j, (n+1)}$, that is a sum over all centers of species $K$
(with $I,J,K$ distinct).  The importance of this general iterative
procedure is that it generates all these terms in the bubble equations
in exactly the correct form and yields the formula (\ref{extraY}) for
a general position, $a_i$, of the probe:
\begin{equation} \label{genY}
 Y^{(J,I)}_{j, i} ~=~ |\epsilon_{IJK}| \sum_k \frac{\left(a_j-a_{i}\right)\left(a_j-a_k\right)}{|a_j-a_{i}||a_j-a_k|} \frac{hqd^{(I)}_{n+1}\hat{k}^{(J)}_j\hat{k}^{(K)}_k}{a_{n+1}a_ja_kV_{n+1}V_jV_k}  \,.
\end{equation}

 \goodbreak
 
\subsection{Topology, charge shifts and back-reacting probes}
 \label{multitubes2}

 To understand the charge shifts and their dependence on the location
 of the supertubes it is important to recall the geometric structure
 underlying the almost-BPS supergravity solutions.  In an appropriate
 duality frame\footnote{This can be realized by performing three
   generalized spectral flows on almost-BPS solutions
   \cite{Dall'Agata:2010dy}.} a solution containing multiple
 supertubes becomes smooth, and all the charges come from fluxes
 wrapping topologicaly non-trivial cycles.  These cycles can be
 described in terms of $U(1)$ fibers over paths that end at the
 locations of the supertubes.  The supertube locations are precisely
 where one of the $U(1)$ fibers pinches off and this, in turn, defines
 cycles and their intersections. The magnetic dipole charges of the
 supertube then correspond to cohomological fluxes through these
 cycles. Hence, in this particular duality frame the back-reaction of
 a supertube corresponds to blowing up a new cycle and replacing a
 singular magnetic source with a cohomological flux.  If one starts
 out with a particular supergravity solution and makes a particular
 choice of homology basis, then introducing a new supertube and
 back-reacting it will involve blowing up a new cycle and perhaps
 pinching off other cycles in order to achieve this. Hence, this will
 involve generically a reshuffling of the homology basis.

 The presence of the Chern-Simons term in the electromagnetic action
 means that the interaction of pairs of magnetic charges can source
 electric charges and so the change of homology basis arising from the
 back-reaction of a supertube can lead to shifts of electric charges.
 Because the magnetic charges on the compact cycles are quantized and
 the magnetic contributions to electric charges are determined though
 the intersection form, one would expect all charge shifts to be
 quantized.  This is, indeed, precisely what one finds in all the BPS
 solutions: the shifts, if they are non-zero, are indeed quantized.
 What distinguishes the non-BPS solutions is that there is a
 non-vanishing, normalizable flux on a non-compact cycle that extends
 to infinity and this flux depends upon the supergravity parameters
 and moduli.  It is the interactions between the fluxes on this
 non-compact cycle that leads to moduli-dependence of the shifts.

 One can recast this geometric picture in more physical terms through
 a careful examination of Dirac strings.  Because a supergravity
 supertube carries a magnetic charge, the supertube comes with Dirac
 strings attached (these are, of course, an artifact of trying to
 write a vector potential for a topologically non-trivial flux).  A
 supertube wraps a $U(1)$ fiber in the background and to define its
 configuration properly one must specify precisely which Dirac strings
 are being wrapped by the supertube.  This defines the dipole-dipole
 interaction between the probe and the background\footnote{This
   wrapping choice determines the new homology element and its
   intersections with other other elements of homology.} and the
 difference between wrapping and not wrapping will appear as a shift
 of the electric charges that it contributes to the supergravity
 solution.

 If the supertubes are all colinear then one can choose all the Dirac
 strings to follow the axis of symmetry out to infinity.  One can then
 set up a configuration in which an outermost probe supertube wraps
 the Dirac strings of all the other supertubes, and hence its charges
 get shifted. On the other hand the tubes at the interior of this
 configuration will not feel the Dirac string of the outermost
 supertube, and hence the relationship between their supergravity
 charge and their quantized charge does not shift.  In the duality
 frame where the multi-supertube solution is smooth, this corresponds
 to blowing up a homology cycle at the outer edge of the original
 configuration.

 If one were to place the probe supertube so that it is the closest to the
 Taub-NUT center, its Dirac string would generically affect all the
 other supertubes, and would change the relation between quantized and
 supergravity charges.  This corresponds to blowing up a different new
 homology element and reshuffling the homology basis. Hence, if one
 keeps the supergravity charges of the back-reacting supertubes fixed,
 as one must do in a probe approximation, bringing a supertube to a
 point that is closer to the center of Taub-NUT than the other supertubes will change the
 quantized charges of these supertubes. Thus, the resulting
 configuration will not have the same quantized charges on the centers
 as when the supertube is at the outermost location, and hence belongs
 to different sector of the multi-centered solutions.  There are, of
 course, similar consequences to bringing the probe supertube to some
 point in the middle of the back-reacted supertube centers.

 One can also take a more pragmatic perspective and try to understand 
 the charge shifts by compactifying  the multi-supertube solutions to obtain 
 a multi-center almost-BPS  solution in four dimensions. The five-dimensional, 
 smooth solutions we  discuss here are singular at the supertube centers in 
 four dimensions, but this does not impede the calculation of the conserved charges 
 at the  singularities. As explained in \cite{Bossard:2012ge}, these conserved
 charges differ in general from the supergravity charge parameters by
 dipole-dipole position-dependent shifts similar to the ones found
 here, and it would be interesting to see whether one can reproduce
 our results using the four-dimensional KK reduction formulae of
 \cite{Bossard:2012ge} (starting, for example, from equation (266) on page
 55).

\subsection{Quantized charges, supergravity parameters and probes}
 \label{Probes}

 As we have seen in the previous section, bringing a supertube to
 some point in the middle of a multi-center supertube solution changes
 the relation between the quantized charges and the supergravity
 parameters of the supertube centers at its exterior. Hence, all the
 probe calculations that describe supertubes at interior locations in
 a multi-center solution are not self-consistent, because the quantized
 charges of the other centers before bringing in the probe are not the
 same as the quantized charges with the probe inside. One can think
 of this as coming from the fact that all probe supertubes come
 with Dirac strings attached, and when these Dirac strings touch the
 other centers, the relation between the supergravity and quantized
 charges of these centers change. Thus, the only probe supertube that
 one can bring without shifting everybody else's quantized charges is
 one that lays at the outermost position and whose Dirac string
 extends away from the supertubes and towards infinity.

 An immediate corollary of this is that if one wants to calculate the
 amplitude for a supertube to tunnel to a vacuum across another
 center, this calculation cannot be done using the DBI action of that
 supertube and treating it as a probe in a fixed supergravity
 background. Instead one would  have to change the solution as the probe
 moves around, and this cannot be done off-shell. One can wonder whether
 there exists {\em any} way to compute this tunneling amplitude. Again, the
 correct probe for this would be a supertube {\it with Dirac strings
 attached}, such that one would change the supergravity charges of the
 back-reacted solution as one moves the probe around, in such a way
 that the quantized charges stay the same. Unfortunately, there is no known
 action for such a probe, so probing the interior of a multi-supertube
 solution with a probe supertube does not correspond to a physical
 process. Hence, the calculation we performed in Section
 \ref{colineargeneral}, where we took the supergravity parameters to
 be fixed and treated the background of $n$ supertubes merely as one
 would treat any other supergravity background ignoring the details of
 how it might have been assembled from other supertubes, should be
 interpreted as a formal calculation, which does not correspond to a
 physical process, but which does however reproduce the bubble
 equations of all the interior supertube centers. It would be clearly
 interesting to understand the reason for this.

It is also important to stress out that this charge shift subtlety  only
affects the validity of the probe calculation when the other centers
are supertubes or black rings, that interact directly with the Dirac
string of the probe. If the other centers are bubbled Gibbons-Hawking centers,
where the geometry is smooth, the presence of a Dirac string does not
change their four-dimensional charge parameters nor the fluxes
wrapping the corresponding cycle in the five-dimensional solution.
Hence, one can use the probe supertube action to calculate tunneling
probabilities of metastable supertubes in bubbling geometries
\cite{Bena:2011fc,Bena:2012zi}. Similarly, the description of
multi-center non-BPS solutions using supergoop methods
\cite{Anninos:2012gk}, being intrinsically non-gravitational, is not
affected by this subtlety.

\section{Conclusions}
 \label{Conclusions}

 We have shown that a brane probe in an almost-BPS background of
 supertubes can capture the complete supergravity data of the
 background in which the probe becomes a fully back-reacted source.  A
 similar result was established for BPS solutions in
 \cite{Bena:2008dw} but it is rather surprising that this can also be
 achieved for almost-BPS solutions in which the supersymmetry is
 broken, albeit in a rather mild manner.  It is surprising because
 going from a probe to a back-reacted supergravity solution represents
 going from weak to strong coupling in the field theory on the brane
 and when supersymmetry is broken one would expect the parameters of
 this field theory to be renormalized.  Our result therefore suggests
 that, even though supersymmetry is broken, the couplings of the field
 theory that govern the probe location are still protected.  This does
 not mean that the theory is unchanged relative to its BPS
 counterpart: there are terms in the probe action that come from the
 supersymmetry breaking and these exactly reproduce the corresponding
 terms found in the supergravity solution.

 We suspect that the non-renormalization of the relevant parts of the
 probe action arises from the rather special form of the supersymmetry
 breaking: the probe is supersymmetric with respect to every other
 center in the supergravity solution taken individually and the
 supersymmetry breaking arises from the fact that these
 supersymmetries are incompatible between multiple centers. Indeed the
 supersymmetry breaking terms depend on the product of the charges and
 dipole charges of three or more centers, and do not have the
 ``two-body'' structure of the terms that appears in the BPS bubble
 equations. It would be interesting to try to extend our analysis to
 the other known class of non-BPS extremal multi-center solutions, the
 interacting non-BPS black holes
 \cite{Bossard:2011kz,Bossard:2012xsa}, and see if the bubble
 equations of these solutions are also not renormalized when one goes
 from weak to strong effective coupling.

 One of the new features of our analysis is that the supergravity
 charge parameters and the quantized charges of the supertubes need to
 be shifted relative to one another in order to match the supergravity
 bubble equations and the probe radius relations.  By taking a
 flat-space limit, we saw that part of this difference was related to
 the choice of how the supertube wraps Dirac strings, or equivalently,
 how the cycle is blown up after back-reaction.  This accounts for a
 quantized shift but in a Taub-NUT background the shift is no longer
 obviously quantized because it depends upon moduli.  Quantized shifts
 have also been encountered and understood in the study of black rings
 \cite{deWit:2009de, Hanaki:2007mb, Bena:2008dw} but it would be very
 nice to understand how to extract the quantized charges, and hence
 the more general moduli-dependent shifts, via a pure supergravity
 calculation.  A good starting point may be to use the
 four-dimensional reduction formulas of \cite{Bossard:2012ge} as well
 as a judicious accounting of the integer charge shifts caused by
 Dirac string, to try to reproduce the shifts we find.

 Our results also suggest further interesting supergravity
 calculations.  For rather mysterious reasons, explicit solutions of
 the almost-BPS equations still elude us for non-axi-symmetric
 configurations. Such solutions are extremely simple in the BPS system
 but for the almost-BPS system even the three-centered,
 non-axi-symmetric solution is not known. The most one could get is an
 implicit solution in terms of integrals that one can evaluate
 asymptotically \cite{Bossard:2012ge}.  Finding the three-centered
 solution is particularly important because one can then use this
 solution and the methods of \cite{Bena:2009en} to generate the
 general, non-axi-symmetric, multi-centered solutions.  A two-centered
 solution is, of course, trivially axi-symmetric and it is easy to
 introduce a probe in any location.  It would therefore be very
 interesting to see if such a probe could be used to gain insight into
 the structure of the full supergravity solution.

 Given that our results suggest that there are non-renormalization
 theorems that protect the bubble equations of almost-BPS multi-center
 solutions, it would be extremely interesting to investigate whether
 almost-BPS solutions could be described in the regime of parameters
 where none of the centers is back-reacted, using a suitable
 generalization of quiver quantum mechanics of \cite{Denef:2002ru}.
 Clearly, this generalization will have to account for the
 supersymmetry-breaking terms, which is highly non-trivial: these terms
 involve three- and four-center interactions, and at first glance no
 theory of strings on multiple D-branes appears capable of producing
 such a term.  It would be very nice to understand how the charge
 shifts and supersymmetry breaking terms emerge from the field theory
 on the branes and whether the effective number of hypermultiplets may be 
 different from the number of hypermultiplets appearing in the
 Lagrangian.  This could be the effect of some of these hypermultiplets becoming
 massive and no longer contributing to the low energy dynamics.
 It is thus not too difficult to imagine that our results might be
 derivable from a field theory analysis and we intend to investigate this
 further in the future.

\bigskip
\leftline{\bf Acknowledgements}
\smallskip
We would like to thank Guillaume Bossard, Sheer El-Showk and Stefanos Katmadas for
valuable discussions.  OV and NPW are grateful to the IPhT, CEA-Saclay
for hospitality while this work was initiated.  The work of IB and AP
was supported in part by the ANR grant 08-JCJC-0001-0, and by the ERC
Starting Independent Researcher Grant 240210-String-QCD-BH. The work
of OV and NPW was supported in part by the DOE grant
DE-FG03-84ER-40168. OV would also like to thank the USC Dana and David
Dornsife College of Letters, Arts and Sciences for support through the
College Doctoral Fellowship and the Gerondelis Foundation for
additional support through a fellowship.


\appendix 
\section{Details of the angular momentum vector}
\label{appendixA}
\renewcommand{\theequation}{A.\arabic{equation}}
\renewcommand{\thetable}{A.\arabic{table}}
\setcounter{equation}{0}

The angular momentum vector, $k$, in the metric (\ref{11Dmetric})  is decomposed as in (\ref{kansatz}) and here we summarize the results of \cite{Vasilakis:2011ki} and give the detailed expressions in our conventions. We also examine the relationship between the BPS and non-BPS descriptions for $h=0$. 

\subsection{The function, $\mu$}

 For BPS solutions, $\mu$ is given by the standard expression in  \cite{Bena:2005va, Berglund:2005vb}:
\begin{equation}
\mu ~=~ \coeff {1}{6} \, V^{-2} \, C_{IJK}\,   K_+^I K_+^J K_+^K ~+~  \coeff{1}{2}  \,V^{-1}  \, K_+^I L_I ~+~  M_+\,, 
\label{muBPS}
\end{equation}
where $M_+$ is harmonic function that we will take to be
\begin{equation}
M_+  ~=~ m^+_\infty ~+~ \frac{m^+_0}{r} ~+~   \sum_{j=1}^3  \, \frac{m^+_j}{r_j} ~+~ m^+_{div} \, r \cos \theta  \,.
\label{Mplusdefn}
\end{equation}
For reasons that will become apparent below, we have added an unphysical harmonic term that diverges at infinity.

For the almost-BPS system one can use the results of  \cite{Bena:2009en} to obtain 
\begin{align}
 V \, \mu ~=~ &  \frac{1}{2}\, \sum_{I=1}^3 \, K^I_-   (V + h(L_I -1) )~-~ \frac{q}{2}\, \sum_{i,j=1 \atop j \ne i}^3 \, k_i^-  Q^{(i)}_{j} \, \frac{r^2 -2 a_i r \cos \theta + a_i a_j}{a_i (a_i - a_j)\, r\,  r_i \, r_j }     \nonumber  \\
 & ~+~ \frac{k_1^- k_2^-  k^-_3}{r_1 r_2 r_3}\, \Big[ h^2 ~+~ \frac{q^2\, r \cos \theta}{a_1 a_2 a_3}~+~  \frac{h\, q \, (r^2 (a_1 +a_2 +a_3)+ a_1 a_2 a_3 ) }{2\, a_1 a_2 a_3\, r} \Big]   ~+~ M_-  \,,   
  \label{munonBPS}
\end{align}
where $M_-$ is another harmonic function which we will take to be:
\begin{equation}
M_-  ~=~ m^-_\infty ~+~ \frac{m^-_0}{r} ~+~   \sum_{j=1}^3  \, \frac{m^-_j}{r_j} ~+~ \frac{m^-_{div}}{r^2} \, \cos \theta  \,.
\label{Mminusdefn}
\end{equation}
Again we have added an unphysical harmonic term but this time it diverges at the origin.

Note that regularity at the supertubes requires the following for both classes of solution:
\begin{equation}
m^\pm_1 ~=~  \frac{Q_1^{(2)} \, Q_1^{(3)}  }{ 2\,k^\pm_1}  \,, \qquad m^\pm_2 ~=~  \frac{Q_2^{(1)} \, Q_2^{(3)} }{2 \,k^\pm_2}  \,, \qquad m^\pm_3 ~=~  \frac{Q_3^{(1)} \, Q_3^{(2)}}{ 2\,k^\pm_3}  \,,   \label{mjres}
\end{equation}
%

\subsection{The one-form, $\omega$}

 Following  \cite{Bena:2005va, Berglund:2005vb}, for $a_i>a_j$ we define:
\begin{equation}
\omega_{ij} ~\equiv~  - {(r^2 \, \sin^2\theta+ (r \cos\theta -a_i + r_i)(r \cos\theta - a_j  - r_j)) \over (a_i - a_j)  \, r_i  \, r_j }   \,,
\label{omegaijdefn}
\end{equation}
 To define the BPS angular momentum vector, $\omega$, it is also convenient to introduce:
\begin{eqnarray}
\Delta_1  & \equiv& 2\, \big(h+\frac{q}{a_1}\big)  m^+_1  ~-~ \frac{\Gamma^+_{12}}{(a_2 - a_1)} ~-~ \frac{\Gamma^+_{13}}{(a_3 - a_1)} ~-~  k^+_1  \,, \label{Deltadefn1}\\ 
\Delta_2 & \equiv& 2\, \big(h+\frac{q}{a_2}\big)  m^+_2  ~+~ \frac{\Gamma^+_{12}}{(a_2 - a_1)} ~-~ \frac{\Gamma^+_{23}}{(a_3 - a_2)} ~-~  k^+_2  \,,\label{Deltadefn2} \\
\Delta_3 & \equiv& 2\, \big(h+\frac{q}{a_3}\big)  m^+_3  ~+~ \frac{\Gamma^+_{13}}{(a_3 - a_1)} ~+~ \frac{\Gamma^+_{23}}{(a_3 - a_2)} ~-~  k^+_3 \label{Deltadefn3} \,,
\end{eqnarray}
Note that the bubble equations (\ref{BPSbubble1})--(\ref{BPSbubble3}) and (\ref{mjres}) imply that $\Delta_j =0$, $j=1,2,3$.  We will not impose these conditions here and we will not remove Dirac strings more generally.

We then find that for the BPS solution, $\omega = \omega_\phi d \phi$, with 
\begin{align}
\omega_\phi  ~=~ &   \frac{1}{2}\,   \big( \Gamma^+_{21} \, \omega_{21}  +  \Gamma^+_{31} \, \omega_{31} +  \Gamma^+_{32} \, \omega_{32}  \big) ~+~   q  \, \sum_{i =1}^3 \, \frac{m_i^+}{a_i\, r_i} 
(r \, \sin^2\theta +(r \cos\theta -a_i + r_i)( \cos\theta   - 1)  \nonumber  \\
 &    ~+~   \frac{1}{2}\,   \sum_{i =1}^3 \,\Delta_i\,  \frac{r \cos\theta  - a_i}{ r_i}~+~ \kappa^+  ~-~  \frac{1}{2}\,   \Big( \frac{\Gamma^+_{12}}{(a_1 - a_2)} + \frac{\Gamma^+_{13}}{(a_1 - a_3)} + \frac{\Gamma^+_{23}}{(a_2 - a_3)}   \Big)  \nonumber  \\
& ~+~  \Big( q\, \sum_{i =1}^3 \frac{m_i^+}{a_i}  - h\, m_0^+  \Big) \,(1-\cos \theta)  - q\, m_\infty^+ \cos \theta ~+~ m^+_{div} (\coeff{1}{2} h r^2 + q r) \sin^2 \theta  \,,   
  \label{omegaBPS}
\end{align}
and where $\kappa^+$ is a constant of integration.

One may use the results of \cite{Bena:2009en} to find $\omega_\phi$ for the almost-BPS solution:
\begin{align}
\omega_\phi ~=~&    -  \frac{1}{2}\,   \ \sum_{i =1}^3 \, \frac{k_i^-}{a_i \, r_i} \big( h\, a_i \,  (r \cos\theta  - a_i) + q \, (r   - a_i \cos\theta)  \big)
 \nonumber \\
 &- \frac{1}{2}\, \sum_{i,j=1 \atop j \ne i}^3  \frac{k_i^-  Q^{(i)}_{j} }{ a_i (a_i - a_j)\, r_i \, r_j }   \big[ h\, a_i \, \big(  r^2 - (a_i +a_j) r \cos \theta + a_i a_j\big) \nonumber \\
 &\qquad \qquad \qquad \qquad \qquad \qquad  \qquad+ q\, \big(  (r^2  + a_i a_j\cos \theta )- r ( a_i \cos 2\theta + a_j)\big) \big]
   \nonumber  \\
    &-\frac{k_1^- k_2^-  k^-_3}{a_1 a_2 a_3 \, r_1 r_2 r_3} \, \big[ \, q^2\, r^2 \sin^2 \theta ~+~ \coeff{1}{2}\,  h\, q \, \big(r^3 -  (a_1 +a_2 +a_3) r^2 \cos \theta   \nonumber \\
 &\qquad \qquad \qquad \qquad \qquad \qquad  \qquad +  (a_1 a_2 +a_1 a_3+a_2 a_3)r   - a_1 a_2 a_3 \cos \theta\big) \big]
   \nonumber  \\
& -  \kappa^- ~-~ \frac{m^-_{div}}{r} \sin^2\theta ~+~ m_0^-  \cos\theta  ~+~   \sum_{i =1}^3 \, \frac{m_i^-}{r_i} 
(r \cos\theta -a_i)   \,,   
  \label{omegaNBPS}
\end{align}
where $\kappa^-$ is another constant of integration.

\subsection{The Minkowski-space limit}

For $h=0$ the BPS and non-BPS solutions must be the same and in Section \ref{MinkLim}  it was argued that this should be achieved by the coordinate change (\ref{BPStoNBPS}). For the angular momentum vector this means
\begin{align}
\mu_- (d\psi_-  -  q\, \cos \theta \, d \phi_-) ~+~ \omega_{\phi \,  - } d \phi_-  & ~=~   \mu_+ (d\psi_+ +q\,  \cos \theta \,d \phi_+) ~+~ \omega_{\phi \, +} d \phi_+    \nonumber \\
 & ~=~   \tilde \mu_+ (- q \, d\phi_-  +  \cos \theta \, d \psi_-) ~+~  \frac{1}{q} \, \tilde \omega_{\phi \, +} d \psi_- 
\label{kequiv}
\end{align}
where the quantities with the tildes have the replacement $k_I^+  \rightarrow \widehat k_I$.  This implies that we must have (for $h=0$):
\begin{equation}
\mu_-  ~=~    \tilde \mu_+  \cos \theta   ~+~    \frac{1}{q} \, \tilde \omega_{\phi \, +}   \,, \qquad  \omega_{\phi \,  - }   ~=~   \tilde \omega_{\phi \, +} \,  \cos \theta  ~-~ q\,  \,  \tilde \mu_+  \sin^2 \theta    \,.
\label{muomrelns}
\end{equation}
We have verified this by explicit computation and it works provided that one also makes the substitutions:
\begin{equation}
m_i^+  ~\rightarrow~   \frac{a_i}{q}  \,  m_i^-  \,, \quad m_0^+  ~\rightarrow~   \frac{1}{q}  \, m_{div}^- \,, \quad m_\infty^+  ~\rightarrow~    \frac{1}{q}  \,  \kappa^-  \,, \quad m_{div}^+  ~\rightarrow~    \frac{1}{q}  \, m_\infty^-  \,, \quad \kappa^+  ~\rightarrow~    m_0^-  \,.
  \label{othersubs}
\end{equation}
Thus the transformation (\ref{BPStoNBPS}) does indeed map the complete BPS description to the almost-BPS description. Note also that we have included the terms $m_{div}^\pm$ so as to complete this dictionary.






\end{document}